\def\G#1{\Gamma\left(#1\right)}
\def\eps{\epsilon}
\def\be{\begin{eqnarray}}
\def\ee{\end{eqnarray}}
\documentstyle[preprint,revtex]{aps}
\begin{document}
\thispagestyle{empty}
\preprint{Alberta Thy-25-92 }
\preprint{August 1992}
\vspace{1in}
\begin{title}
On the QCD corrections to the charged Higgs decay\\ of  a  heavy
quark.
\end{title}
\hspace{1.5in}
\author{\begin{center}Andrzej Czarnecki\cite{ac} and Sacha Davidson
\cite{sd} \end{center}}
\vspace{-1cm}
\begin{instit}\begin{center}
 Department of Physics, University of Alberta, Edmonton, Canada T6G
2J1
\end{center}\end{instit}
\hspace{3in}
\begin{abstract}
Using dimensional regularization for both infrared and ultraviolet
divergences, we confirm that the QCD corrections to the decay width
$\Gamma(t\to H^+b)$ are equal  to  those to $\Gamma(t\to W^+b)$ in
the limit of a large $t$ quark mass.

\end{abstract}

\narrowtext
\newpage
Many extensions of the Standard Model contain more than the one
necessary Higgs
doublet, and the new degrees of freedom appear as extra Higgs
scalars, some of which are charged (see \cite{liyuan90} and
references therein). We consider here the decay  $t\to H^+b$ in a
model with two Higgs doublets. The first order  QCD correction to
this process has been calculated by two groups
\cite{liyuan90,liyuan92,liuyao92} who disagree with each other. In
this letter we present a calculation of this
correction in the limit of a very heavy top quark, i.e., neglecting
the masses of the bottom quark and the Higgs boson. The effect of
finite $m_H$ and $m_b$ will be addressed in a forthcoming paper. We
use dimensional regularization to cope with both the ultraviolet and
infrared divergences, which greatly simplifies the calculation,
especially  the real radiation part.

We take $H_1$ and $H_2$ to be the doublets whose vacuum expectation
values respectively give masses to the down and up type quarks. The
physical charged Higgs $H^+$ is a linear combination of the charged
components of $H_1$ and $H_2$, so if we neglect all the Yukawa
couplings except that of  $H_2$ to the third generation
$=h^{(2)}_{tt}$, the top only couples to the  $H_2$ component of
$H^+$. The interaction Lagrangian relevant to the
decay $t\to H^+b$ is then:

\be
{\cal L}&=&h^{(2)}_{tt}\cos\beta V_{tb} H^+\bar t
\left({1-\gamma_5\over 2}\right)b+h.c. \nonumber \\
&=&{g\over 2\sqrt{2}m_W}V_{tb}\cot\beta m_tH^+\bar t
\left(1-\gamma_5\right)b+h.c.
\ee
where $H^+=\cos\beta H_2^+-\sin\beta H_1^+$ and
$\cot\beta=\left<H_1\right>/\left<H_2\right>$ is the ratio of vacuum
expectation
values of the two Higgs doublets.

In the following calculations we take the space-time dimension to
be $D=4-2\eps$. The mass of the decaying quark is taken to be the
renormalization mass scale, and we also use it as a unit of energy:
$m_t=1$. In the limit of a very heavy top quark the above
interaction leads to the tree-level decay rate:
\be
\Gamma^{(0)}\left(t\to H^+b\right)={G_F\over \sqrt{2}}
{\G{1-\eps}\over 2^{3-2\eps}\pi^{1-\eps}\G{2-2\eps}}
\cot^2\beta \left| V_{tb} \right|^2
\to {G_F\over 8\sqrt{2}\pi}
\cot^2\beta \left| V_{tb} \right|^2.
\ee

The first order QCD corrections to this formula arise due to

virtual gluon exchange and  radiation of a real gluon. We first
deal with the virtual gluon correction to the vertex $tH^+b$. In
the limit $m_b=m_H=0$ the spinor structure of this vertex remains
unchanged and the unrenormalized correction amounts to the
multiplication of the tree level rate by a factor
\be
\Lambda=C_Fg_s^2{\G{1+\eps}\over (4\pi)^{D\over 2}}\left(-{1\over
\eps^2}+{2\over\eps}\right).
\ee
where the colour factor $ C_F$ is  $4/3$ for $SU(3)$. The counterterm
for this vertex involves the wave function and mass renormalization
constants \cite{bl80,liyuan90}:
\be
\Lambda_{c.t.}={1\over 2}\left(Z_2^t-1\right)+{1\over
2}\left(Z_2^b-1\right)
 -{\delta m_t\over m_t}
 -{\delta m_b\over m_b}.
 \ee

 If we use the same $\eps$ to regularize both UV and IR divergencies
we obtain for the renormalization constants:
\be
 Z_2^t-1&=& -{\delta m_t\over m_t}=C_Fg_s^2{\G{1+\eps}\over
(4\pi)^{D\over 2}}\left(-{3\over
\eps}-4\right),\nonumber\\
Z_2^b-1&=& {\delta m_b\over m_b}=0.
\ee
The contribution of the virtual correction to the decay rate is:
\be
\Gamma^{(1)}_{virt}\left(t\to
H^+b\right)&=&2\left(\Lambda+\Lambda_{c.t.}\right)
\Gamma^{(0)}\left(t\to H^+b\right)\nonumber\\
&=&{G_F\over \sqrt{2}}\cot^2\beta \left| V_{tb} \right|^2\alpha_s
{2^{-3+4\eps}\pi^{-2+2\eps}\over 3\G{2-2\eps}}
\left(-{2\over \eps^2}-{5\over \eps}-12-{\pi^2\over 3}\right).
\ee

We now turn our attention to the effect of real gluon radiation from
the initial or final quark. If we denote the amplitudes for these
processes by

${\cal A}_1$ and ${\cal A}_2$ respectively, the contribution of the
real radiation to the decay width is:
\be
\Gamma^{(1)}_{real}\left(t\to H^+bG\right)= {1\over 2}{G_F\over
\sqrt{2}}\cot^2\beta \left| V_{tb} \right|^2   4\pi \alpha_s \int
dR_3\left(t;b,H,G\right) \left| {\cal A}_1+{\cal A}_2\right|^2
\ee
where the coupling constants have been factored out and $t,b,H$ and
$G$ denote the four-momenta of the initial and final quarks, charged
Higgs boson and the gluon.

 The advantage of using  dimensional regularization for the infrared
and   colinear divergences is that we need not introduce a  mass for
the gluon and the integration over three body massless phase space is
very simple. We choose to parametrise it by the variables $x=2t\cdot
G$ and $z=1-2t\cdot b$ in terms of which the three body phase space
integration becomes:
\be
\int dR_3\left(t;b,H,G\right) ={2^{4\eps-7}\pi^{2\eps-3}\over
\G{2-2\eps}}
\int_0^1 {dx \over \left(1-x\right)^\eps}
\int_0^x {dz \over z^\eps\left( x-z\right)^\eps}.
\ee
After summing over the polarizations of the $b$ quark and the gluon,
and averaging over the polarizations of the $t$ quark, the squares of
the amplitudes become:
\be
\left| {\cal A}_1\right|^2 &=& -{4\over x^2}
\left[2(1-x)+x(1-\eps)(z-x) \right] \nonumber \\
\left| {\cal A}_2\right|^2 &=& {4x\over x-z} (1-\eps) \nonumber \\
{\cal A}_1{\cal A}_2^*+{\cal A}_2{\cal A}_1^* &=& {8\over x(x-z)}
\left[1-x+x(1-\eps)(z-x) \right]
\ee
The integration over the phase space can be done exactly in any
dimension. In the limit $\eps \to 0$ the contribution of the real
radiation becomes:
\be
\Gamma^{(1)}_{real}\left(t\to H^+b\right)={G_F\over
\sqrt{2}}\cot^2\beta \left| V_{tb} \right|^2\alpha_s
{2^{-3+4\eps}\pi^{-2+2\eps}\over 3\G{2-2\eps}}
\left({2\over \eps^2}+{5\over \eps}+17-{\pi^2}\right).
\ee

Although the respective phase space integrations of

$\left| {\cal A}_1\right|^2, \left| {\cal A}_1\right|^2$ and

${\cal A}_1{\cal A}_2^*+{\cal A}_2{\cal A}_1^*$  give different
results from the analogous amplitudes  with a $W^+$ replacing the
charged scalar,   their sum nevertheless gives the same total
contribution in both processes. This is in agreement with the general
argument based on the equivalence theorem in ref.~\cite{liuyao92}.

Finally we add the effects of the virtual and real gluons to obtain
the first order QCD correction, so  the decay rate (with $m_t$
reinstated) becomes:
\be
\Gamma\left(t\to H^+b\right)={G_F\over 8\sqrt{2}\pi}m_t^3
\cot^2\beta \left| V_{tb} \right|^2\left[1+{\alpha_s\over
3\pi}\left(5-{4\pi^2 \over 3}\right)\right].
\ee

This is identical to the result obtained in ref.~\cite{liuyao92} and
also to the analogous correction to the decay $t\to W^+b$
\cite{jk1,cza90,liuyao91,oakes91}. If we take $\alpha_s=0.1$ the
first order correction in the limit $m_b=m_H=0$ is approximately
equal $-0.87\%$. This is in disagreement with the value reported in
ref.~\cite{liyuan90,liyuan92}.

\acknowledgments

We   gratefully acknowledge support for our doctoral scholarships:
A.C.~from the Killam Foundation and S.D.~from NSERC. This research
was also partially supported by a grant to A.~N.~Kamal from the
Natural Sciences and Engineering Research Council (NSERC) of Canada.

 This preprint was typeset using $REVTEX$.


\begin{references}
\bibitem[*]{ac}Bitnet address: USERAPRU@UALTAMTS
\bibitem[**]{sd}Permanent address after October 1st,
1992: Center for Particle Astrophysics, \\University of California
at Berkeley, Berkeley, California 94720.
\bibitem{liyuan90}
C. S. Li and T. C. Yuan, Phys.~Rev. D42 (1990) 3088.
\bibitem{liyuan92}
C. S. Li et al., Phys.~Lett. B285 (1992) 137.
\bibitem{liuyao92}
J.~Liu and Y.~P.~Yao, preprint UPR-0508T (1992).
\bibitem{bl80}
E.~Braaten and  J.P.~Leveille, Phys.~Rev. D22 (1980) 715.
\bibitem{jk1}
 M. Je\.zabek and J. H. K\"uhn, Nucl. Phys. B314
(1989) 1.
\bibitem{cza90}
A.~Czarnecki, Phys.~Lett.~B252 (1990) 467.
\bibitem{liuyao91}
J.~Liu and Y.~P.~Yao, Int.~J.~Mod.~Phys. A6 (1991) 4925.
\bibitem{oakes91}
C.~S.~Li, R.~J.~Oakes and T.~C.~Yuan, Phys.~Rev.~D43 (1991) 3759.

\end{references}
\end{document}